\documentstyle[twocolumn]{jpsj}

\def\ggs{\buildrel\textstyle > \over {\hbox{\raise0.2ex\hbox{$\sim$}}}}
\def\lls{\buildrel\textstyle < \over {\hbox{\raise0.2ex\hbox{$\sim$}}}}
\def\gsim{\,\lower0.75ex\hbox{$\ggs$}\,}
\def\lsim{\,\lower0.75ex\hbox{$\lls$}\,}

\newcommand{\Js}{J_{\rm s}}

\newcommand{\kb}{{\bf k}}
\newcommand{\qb}{{\bf q}}

\newcommand{\simg}{\stackrel{>}{_\sim}}
\newcommand{\siml}{\stackrel{<}{_\sim}}

\def\runtitle{Phase Diagram of the Electron-Doped Cuprate Superconductors}
\def\runauthor{Akito {\sc Kobayashi} {\it et al.}}

\title
{Phase Diagram of the Electron-Doped Cuprate Superconductors}

\author
{
Akito {\sc Kobayashi}, Atsushi {\sc Tsuruta}, 
Tamifusa {\sc Matsuura} and Yoshihiro {\sc Kuroda}
}

\inst
{
Department of Physics, Nagoya University, Nagoya 464-8602 \\
}

\abst{
We investigate the $T$-$\delta$ phase diagram of the electron-doped systems in high-$T_{\rm c}$ cuprates where $\delta$ is doping rate.
Taking the ingap state and the super-exchange interaction $J_{\rm s}$, we calculate the superconducting transition temperature $T_{\rm c}$, the antiferromagnetic transition temperature $T_{\rm N}$, the NMR relaxation rate $1/T_1$ with the antiferromagnetic fluctuations in the fluctuation-exchange (FLEX) approximation and with the superconducting fluctuations in the self-consistent $t$-matrix approximation.
Obtained phase diagram has common features as those in the hole-doped systems, including the antiferromagnetic state, the superconducting state and the spin gap phenomenon.
Doping-dependences of $T_{\rm N}$, $T_{\rm C}$ and $T_{\rm SG}$ (spin gap temperature) are, however, different with those in the hole-doped systems.
These differences are due to the intrinsic nature of the ingap states which are intimately related with the Zhang-Rice singlets in the hole-doped systems and are correlated $d$-electrons in the electron-doped systems, respectively, which has been shown in the $d$--$p$ model.
}

\kword{electron-doped cuprate, pseudogap, spin gap, superconductivity, superconducting fluctuation, spin fluctuation, superexchange interaction, d-p model} 

\begin{document}
\input epsf

\sloppy
\maketitle

In the high-$T_{\rm c}$ cuprate superconductors, the superconducting (SC) states appear as holes or electrons are doped into the antiferromagnetic (AF) insulator\cite{Sato,Takagi}.
The SC states in the electron-doped cuprates have $d_{x^2 -y^2}$-symmetry as those in the hole-doped cuprates do\cite{Tsuei,Armitage,TSato}, which suggests existence of a common mechanism in both kinds of cuprates.
The doping ranges in which the AF states and the SC states exist in the electron-doped cuprates are, however, defferent from those in the hole-doped cuprates.
In the hole-doped cuprates, the AF states exist in narrow doping range ($0 \le \delta \siml 0.02$), and the SC states exist in wide doping range ($0.05 \siml \delta \siml 0.25$).
In the electron-doped cuprates, on the other hand, the AF states exist in wider range ($-0.13 \siml \delta \le 0$) and the SC states appear in narrower range ($-0.18 \siml \delta \siml -0.13$) than those in the hole-doped cuprates.

In our previous study we have obtained the $T-\delta$ phase diagram in the hole-doped region with the AF fluctuations in the fluctuation-exchange (FLEX) approximation and the SC fluctuations in the self-consistent $t$-matrix approximation\cite{pgafl}, which have been based on the ingap state and the super-exchange interaction $J_{\rm s}$ obtained in the $d$--$p$ model by using the slave boson technique and the $1/N$-expansion theory\cite{Jichu,Ono2,Fukagawa}.
We have obtained the phase diagram including the AF state, the SC state with $d_{x^2 -y^2}$-symmetry and the spin gap phenomenon induced by the SC fluctuation\cite{pgdp1,pgsc,Kosuge}, which are consistent with those observed in the hole-doped cuprates\cite{Sato,Zheng}.

In the present study we investigate the $T-\delta$ phase diagram in the electron-doped region by calculating the superconducting transition temperature $T_{\rm c}$, the antiferromagnetic transition temperature $T_{\rm N}$ and $1/T_1T$ of NMR by using the same approximation as that taken in our previous study\cite{pgafl}.
Obtained phase diagram has common features as those in the hole-doped region, including the AF state, the SC state and the spin gap phenomenon.
Doping-dependences of $T_{\rm N}$, $T_{\rm C}$ and $T_{\rm SG}$ (spin gap temperature) are, however, different from those in the hole-doped systems.
The AF state exists approximately in 3 times wider range than those obtained in the hole-doped region.
The SC state, on the other hand, appears in narrower range.
In addition we show that the spin gap phenomenon appears in narrower range than those obtained in the hole-doped region.
The electron-hole asymmetry is due to the intrinsic nature of the in-gap state described in the followings, while the electron-hole asymmetry has been studied by single band models such as the $t$--$J$ model or the single band Hubbard model by introducing the long range hoppings\cite{Suzumura,Tohyama,Kuroki,Kontani,Kondo,Manske,Yanase}.

We take the $d$--$p$ model which is a kind of the two-band Hubbard model for describing the electronic system in the ${\rm CuO}_2$ plane:
\begin{eqnarray} 
H &=& \varepsilon_p \sum_{\kb \sigma}c_{\kb \sigma}^+c_{\kb \sigma} 
       +  \varepsilon_d \sum_{i \sigma}d_{i\sigma}^+d_{i\sigma}
\nonumber \\       
&+& N_{\rm L}^{-\frac12}\sum_{i \kb \sigma}\{t_{i\kb}
           c_{\kb \sigma}^+d_{i\sigma}b_i^+ + h.c.\}, \label{hamiltonian}
\end{eqnarray}
which is treated within the physical subspace where local constraints
$\hat{Q}_i = \sum_\sigma d_{i\sigma}^+d_{i\sigma} + b_i^+b_i = 1$
strictly hold in order to exclude double occupancy of d-holes\cite{Coleman,Jin}.
In the above, $c_{\kb \sigma}$, $d_{i\sigma}$ and $b_i$ are annihilation operators for a $p$-hole with a wave vector ${\bf k}$ and spin $\sigma$, a pseudo fermion representing a single occupation of $d$-hole of $i$-th site and a slave boson representing a vacancy of the $d$-hole of $i$-th site, respectively, and $t_{i\kb} = t_\kb \exp(-i\kb\cdot{\bf R}_i)$, with
$t_{\kb} = 2t_{dp}[1-\frac12 (\cos k_{x}a + \cos k_{y}a)]^\frac12$
, where $a$ is the lattice constant (we set $a=1$), and $N_{\rm L}$ is the total number of lattice sites.
Here $\varepsilon_p$ and $\varepsilon_d$ stand for the energies of the $p$- and $d$-levels measured from the chemical potential $\mu$, respectively, and $t_{dp}$ stands for the $d$--$p$ transfer.


The $1/N$-expansion is a suitable method for treating correlations systematically.
First, the strong local correlations are suitably included in the self-energy in the leading order of $1/N$-expansion.
The quasi-particle Green's functions of the leading order in the $1/N$-expansion are given by~ \cite{Jichu,Ono2}
\begin{eqnarray} \label{GK} 
G_0 (\kb, \omega ) = \sum_{\gamma=\pm}A_\gamma (\kb )/(\omega  - E_\gamma (\kb )+{\rm i}0^+ ),
\end{eqnarray}
with
$E_\gamma (\kb )=\frac12 [\varepsilon_{p}+\omega_0+\gamma ((\varepsilon_{p}-\omega_0)^2+4b t_{\bf k}^2 )^{1/2} ]$
,
$A_\gamma (\kb )=\gamma (E_\gamma (\kb ) - \omega_0)/(E_+ (\kb )- E_- (\kb ))$,
where $\gamma =-$ and $+$ denote the ingap state and the $p$-band, and $N$ represents the degeneracy of $d$-hole ($N=2$ in the present case).
The binding energy $\omega_0$ and the number $b$ of the slave boson are self-consistently determined together with $\mu$, and then $\mu$ is located in the ingap state.

The system is the charge transfer (CT) type of the Mott insulator at hole doping rate $\delta =0$ and in the region of $\Delta >\Delta_{\rm c}$\cite{Hirashima1}, where $\Delta =\varepsilon_{p} -\varepsilon_{d}$ stands for the CT gap, and the total hole number is given by $n=1+\delta$.
In the hole-doped region, doped $p$-holes form the quasi-particle band called as the ingap state inside the CT gap near the $p$-band.
It is due to the resonating state of the slave boson, which corresponds to the Zhang-Rice singlet\cite{Zhang}.
Those results are consistent with the results of optical experiments\cite{Uchida,Chen,Brookes}.

In the electron-doped region where vacancies of $d$-holes are doped, on the other hand, the ingap state is the $d$-like band inside the CT gap near the localized $d$-state.
The band width and the intensity of the in-gap state approximately equal $\omega_0$ and $b$, respectively, and are proportional to $\vert \delta \vert$ when $\vert \delta \vert \ll 1$.
The intensity in the electron-doped region is almost 3 times larger than that in the hole-doped region, ${\it i. e.}$ the {\it intrinsic} electron-hole asymmetry induced by the local correlation on the $d$-sites.
Those basic features are not changed by taking the higher order terms in the $1/N$-expansion\cite{Tsuruta}.
Those results are consistent with the results obtained recently by the dynamical mean field theory (DMFT) in the two-band Hubbard model\cite{Mutou,Ohashi1,Ohashi3}.




Next, concerning the intersite correlations, it has been shown that the super-exchange interaction $J_{\rm s}$ is the most dominant term within $O(1/N)$ in the vicinity of the CT insulator\cite{Fukagawa}.
We derive the coupled equations of the AF fluctuations and the SC fluctuations which are induced in the ingap state by $J_{\rm s}$.
The AF fluctuations via $J_{\rm s}$ are given by
\begin{equation}
V({\bf q}, \omega ) = J_{\rm s}({\bf q})/(1-J_{\rm s}({\bf q}) \widetilde{\chi}^{\rm s} ({\bf q}, \omega )),
\end{equation}
with
$J_{\rm s}({\bf q})=-J_{\rm s} (\cos (q_x )+\cos (q_y ))$
and the irreducible spin susceptibility
\begin{eqnarray}
\widetilde{\chi}^{\rm s} ({\bf q},\omega ) &=& N_{\rm L}^{-1}\sum_{\bf k} \Omega_{{\bf k}+{\bf q},{\bf k}} \pi^{-1}\int {\rm d}x f(x) \nonumber \\
&\times& [{\rm Im}G({\bf k}+{\bf q},x)G({\bf k},-\omega +x)^* \nonumber \\
&+& G({\bf k}+{\bf q},\omega +x){\rm Im}G ({\bf k},x)],
\end{eqnarray}
where
$\Omega_{{\bf k},{\bf k}^\prime}=b^2 t_{\bf k}^2 t_{{\bf k}^\prime}^2/((E_- ({\bf k}) -\omega_0 )^2 (E_- ({\bf k}^\prime ) -\omega_0 )^2)$ and $f(E) = (\exp\{E/T\} + 1 )^{-1}$.

It was shown that a component with the $d_{x^2 -y^2}$ symmetry among various components of the spin-fluctuation-mediated interaction contributes dominantly to the pairing interaction.\cite{Azami3}
We take only the component with the $d_{x^2 -y^2}$ symmetry of the pairing interaction as
\begin{equation}
v_d =N_{\rm L}^{-2} \sum_{\bf k} \sum_{\bf k^\prime} \psi_{\bf k} V({\bf k}-{\bf k^\prime}, 0) \psi_{\bf k^\prime} ,
\end{equation}
with
$\psi_{\bf k} =\cos (k_x a)-\cos (k_y a)$.

The irreducible pairing susceptibility $\widetilde{\chi}^{\rm p} ({\bf q}, \omega )$ is given by
\begin{eqnarray}
&\widetilde{\chi}&^{\rm p} ({\bf q},\omega )=-N_{\rm L}^{-1}\sum_{\bf k}\psi_{\bf k}^2 \Lambda_{{\bf k},{\bf q}-{\bf k}} \pi^{-1} \int {\rm d}x  \nonumber \\
&\times& [ f(x){\rm Im} G ({\bf k},x) G ({\bf q}-{\bf k},\omega -x) \nonumber \\
&-&f(-x) G ({\bf k},\omega -x) {\rm Im} G ({\bf q}-{\bf k},x)]
\end{eqnarray}
and
$\Lambda_{{\bf k},{\bf k}^\prime}=b^2 t_{\bf k}^2 t_{{\bf k}^\prime}^2/((\pi T)^2 +\omega_0^2 )^2$.
In the above equations, both $\Omega_{{\bf k},{\bf k}^\prime}$ and $\Lambda_{{\bf k},{\bf k}^\prime}$ are the vertices connecting $p$- and $d$-holes.\cite{Fukagawa}

Self-energy corrections due to the SC fluctuations in the $t$-matrix approximation are given by
\begin{eqnarray}
\Sigma_{\rm SCf} ({\bf k},\omega )=\psi_{\bf k}^2 N_{\rm L}^{-1}\sum_{\bf q} \Lambda_{{\bf k},{\bf q}-{\bf k}} \pi^{-1} \int {\rm d}x \nonumber \\
\times [ f(x) T({\bf q},x+\omega ) {\rm Im} G ({\bf q}-{\bf k},x) \nonumber \\
 - g(x) {\rm Im} T({\bf q},x) G ({\bf q}-{\bf k}, x-\omega )^\ast ],
\end{eqnarray}
with the $t$-matrix of the SC fluctuations
\begin{eqnarray}
T({\bf q},\omega )=v_d^2 \widetilde{\chi}^{\rm p} ({\bf q},\omega )/(1-v_d \widetilde{\chi}^{\rm p} ({\bf q},\omega )),
\end{eqnarray}
where $g(E) = (\exp\{E/T\} - 1 )^{-1}$.

Self-energy corrections due to the AF fluctuations in the fluctuation exchange (FLEX) approximation are given by
\begin{eqnarray}
\Sigma_{\rm AFf} ({\bf k},\omega )=N_{\rm L}^{-1}\sum_{\bf q} \Lambda_{{\bf k},{\bf k}-{\bf q}} \pi^{-1} \int {\rm d}x \nonumber \\
\times [ g(x) {\rm Im} V({\bf q},x )  G({\bf k}-{\bf q}, \omega -x) \nonumber \\
 - f(-x) V({\bf q},\omega -x) {\rm Im} G ({\bf k}-{\bf q}, x ) ].
\end{eqnarray}

The renormalized Green's function is given by
\begin{eqnarray}
G ({\bf k},\omega )=[G_{\rm 0} ({\bf k},\omega )^{-1} \nonumber \\
-\Sigma_{\rm SCf} ({\bf k},\omega ) -\Sigma_{\rm AFf} ({\bf k},\omega )]^{-1},
\end{eqnarray}
Both the SC and AF fluctuations are self-consistently treated by solving the coupled equations (3)-(10).

Using the solutions of the above coupled equations, we calculate the NMR relaxation rate, 
\begin{equation}
1/T_1 T =N_{\rm L}^{-1} \sum_q F_{ab}({\bf q}) \lim_{\omega \rightarrow 0} {\rm Im} \chi^{\rm s} ({\bf q}, \omega )/\omega,
\end{equation}
where
$\chi^{\rm s} ({\bf q}, \omega ) = \widetilde{\chi}^{\rm s} (\qb, \omega)/(1-J_{\rm s} ({\bf q}) \widetilde{\chi}^{\rm s} (\qb, \omega))$
and $F_{ab}({\bf q})$ are the form factors\cite{Imai}.
In the cuprates, $1/T_1 T$ indicates the degree of the AF fluctuation.

We define $T_{\rm c}$ as the temperature determined by the condition,
\begin{eqnarray}
\tau =1-v_d \widetilde{\chi}^{\rm p} ({\bf 0},0) - \alpha =0
\end{eqnarray}
, where $\alpha$ represents the degree of the $3$-dimension ($3$D) effect in the quasi-$2$D electronic system.
In the present model, $1-v_d \widetilde{\chi}^{\rm p} ({\bf 0},0)$ does not reach zero at finite temperatures by the effect of $\Sigma_{\rm SCf} ({\bf k},\omega )$, because we take a 2D electronic system in the calculation of the above coupled equations.
We take $\alpha =0.5$ in the present study so as to have finite $T_{\rm c}$ of $O(100{\rm K})$.

We define $T_{\rm N}$ as the temperature determined by the condition,
\begin{eqnarray}
\zeta =1-J_{\rm s}({\bf Q}) \widetilde{\chi}^{\rm s} ({\bf Q},0)-\beta =0
\end{eqnarray}
, where ${\bf Q}=(\pi ,\pi )$ and $\beta$ represents the degree of the $3$D effect.
In the present model $1-J_{\rm s}({\bf Q}) \widetilde{\chi}^{\rm s} ({\bf Q},0)$ also does not reach zero at finite temperatures by the effect of $\Sigma_{\rm AFf} ({\bf k},\omega )$ in the 2D electronic ststem.
We take $\beta =0.02$ in the present study.

We define $T_{\rm 0}$ as the temperature at which the AF fluctuations begin to be appreciable ($\frac{\partial}{\partial T} (1/T_1 T) \vert_{T=T_{\rm 0}} =-50(1/T_1 T) \vert_{T=0.08, \delta =0.1}$ in the present study).
It is assumed that $T_0$ is the upper boundary of the anomalous metallic phase in the hole-doped cuprate\cite{Sato,Kontani,Nishikawa}.

In actual numerical calculations, throughout the present study, we set $2t_{dp} = 1.0$ (eV), $\Delta = 2.5$ and $\Js = 0.1$.
For example, we have $\omega_0 =0.0780$ and $b=0.0515$ at $\delta =0.1$.
The total number of discrete points taken for the $\qb$-summation over the 2D first Brillouin zone is 32 $\times$ 32.
The $\omega$-integral over the region from $-2\omega_0$ to $2\omega_0$ is replaced by the $\omega$-summation of $80$ discrete points.

Figure 1 shows the $T$-$\delta$ phase diagram obtained by solving the above coupled equations, where the phase diagram in the hole-doped region ($\delta >0$) has been obtained in our previous study\cite{pgafl}.
The AF state in the electron-doped region persists to higher doping rate $\delta \cong -0.13$ than the doping rate $\delta \cong 0.05$ where that in the hole-doped region persists.
The SC state appears in narrower doping range $-0.19 \siml \delta \siml -0.13$ than the doping range where that in the hole-doped region appears.
In addition it is shown that the spin gap appears in the doping range $-0.16 \siml \delta \siml -0.13$, which is consistent with the results observed by NMR very recently\cite{Kitaoka}.
Those features account for the phase diagrams observed both in the electron-doped cuprates and in the hole-doped cuprates.

The AF fluctuations dominate in the region of $T_{\rm SG} \siml T \siml T_0$, while the SC fluctuations dominate in the region of $T_{\rm c} < T \siml T_{\rm SG}$\cite {pgafl}.
This crossover behavior induces the anomalous $T$-dependence of $1/T_1T$ as shown in Fig. 2.
As a result of the domination by the SC fluctuations, $T_{\rm c}$ has a maximum at $\delta \cong 0.1$ and decreases with $\delta$ decreasing in the hole-doped region.
If the AF state is removed, $T_{\rm c}$ in the electron-doped region has a maximum at $\delta \cong -0.1$ and decreases as $\vert \delta \vert$ decreases.


\begin{figure}
\def\epsfsize#1#2{0.4#1}
\centerline{
\epsfbox{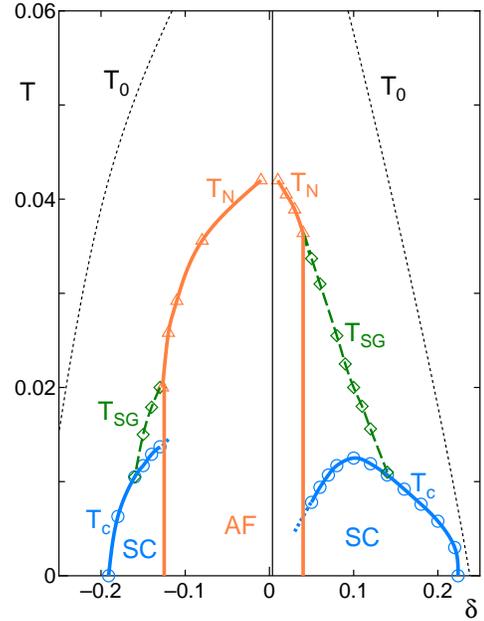}
}
\caption{
The $T$-$\delta$ phase diagram where the system is in the superconducting state with $d_{x^2 -y^2}$ symmetry in the region of $T \le T_{\rm c}$, and is in the antiferromagnetic state in the region of $T \le T_{\rm N}$.
The spin gap temperature $T_{\rm SG}$ is defined as the temperature at which $1/T_1T$ has a maximum as shown in Fig. 2.
The metallic region in $T \siml T_{\rm 0}$ is the anomalous metallic phase.
}
\label{fig.1}
\end{figure}


Figure 2 shows $T$-dependences of $1/T_1T$.
At $\delta \simg -0.12$, $1/T_1T$ increase as $T$ decreases until the system reaches $T=T_{\rm N}$.
At $\delta =-0.13 \sim -0.16$, $1/T_1T$ has a maximum at the spin gap temperature $T_{\rm SG}$ and decreases as $T$ decreases in the region of $T_{\rm c} < T < T_{\rm SG}$, which is the spin gap phenomenon induced by the SC fluctuation.
If the AF state were to be removed, $1/T_1T$ has a maximum at $T_{\rm SG}$ in the region of $\delta \simg -0.12$, and then $T_{\rm SG}$ continues to increase with $\vert \delta \vert$ decreases.

\begin{figure}
\def\epsfsize#1#2{0.4#1}
\centerline{
\epsfbox{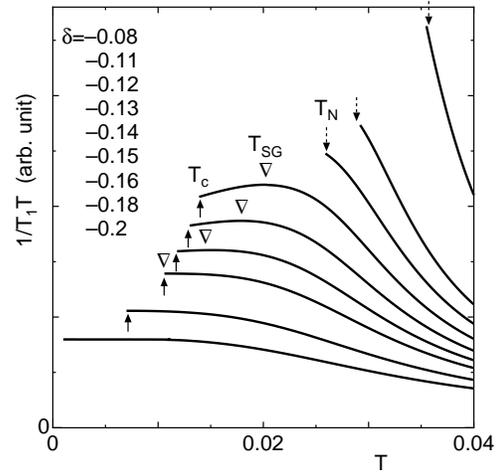}
}
\caption{$T$-dependences of $1/T_1T$ at $\delta =-0.08$, $-0.11$, $-0.12$, $-0.13$, $-0.14$, $-0.15$, $-0.16$, $-0.18$ and $-0.2$.
The triangles represent the spin gap temperature $T_{\rm SG}$ at which $1/T_1T$ has a maximum.
The solid arrows and the dotted arrows represent $T_{\rm c}$ and $T_{\rm N}$, respectively.
}
\label{fig.2}
\end{figure}

Figure 3 shows $T$-dependences of $\tau$ defined in eq. (12) (the solid lines) and $\zeta$ defined in eq. (13) (the dotted lines).
At $\delta =-0.13 \sim -0.18$, $\tau$ decreases as $T$ decreases and reaches zero at $T=T_{\rm c}$.
At $\delta \simg -0.12$, $\zeta$ decreases as $T$ decreases and reaches zero at $T=T_{\rm N}$, while $\zeta$ does not reach zero in the region of $\delta \le -0.13$.
The upturn at $\delta =-0.13$ and the saturation at $\delta =-0.14 \sim -0.16$ in the $T$-dependences of $\zeta$ are mainly due to the SC fluctuations, while those at $\delta \siml -0.16$ are due to the decrease of the AF fluctuation.

\begin{figure}
\def\epsfsize#1#2{0.4#1}
\centerline{
\epsfbox{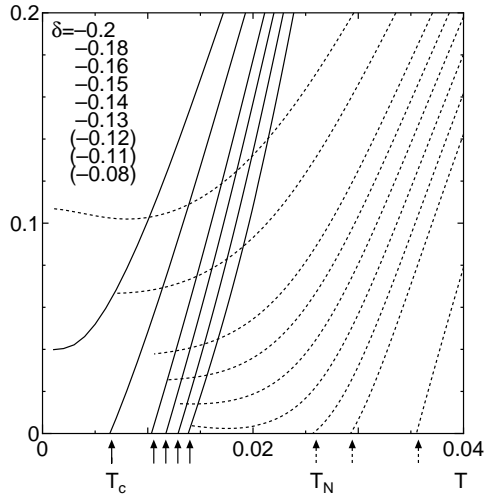}
}
\caption{
The solid lines represent $T$-dependences of $\tau$ at $\delta =-0.2$, $-0.18$, $-0.16$, $-0.15$, $-0.14$ and $-0.13$, which reach zero at $T=T_{\rm c}$.
The dotted lines represent $T$-dependences of $\zeta$ at $\delta =-0.2$, $-0.18$, $-0.16$, $-0.15$, $-0.14$, $-0.13$, $-0.12$, $-0.11$ and $-0.08$, which reach zero at $T=T_{\rm N}$.
}
\label{fig.3}
\end{figure}

In the phase diagram obtained in the present study, $T_{\rm c}$ and $T_{\rm N}$ depend on the phenomenological parameters $\alpha$ and $\beta$, respectively, representing the degree of the 3D effect depending materials.
The qualitative features of the $\delta$-dependences of $T_{\rm c}$ and $T_{\rm N}$ are, however, independent of the choice of these parameters, and $T_{\rm SG}$ is determined without indefiniteness.
By solving the coupled equations (3)-(10) for the anisotropic 3D system, ambiguities due to these parameters should be excluded.

In summary, we have obtained a unified phase diagram accounting for the phase diagrams observed both in the electron-doped cuprates and in the hole-doped cuprates including the AF states, the SC states and the spin gap phenomena.
The AF fluctuations and the SC fluctuations induced in the ingap states by $J_{\rm s}$ are the key factors for describing those, especially the spin gap phenomena.
The electron-hole asymmetry on the $\delta$-dependences of $T_{\rm N}$, $T_{\rm c}$ and $T_{\rm SG}$, is the result from the differences of intrinsic nature of the ingap states due to the strong correlation on the $d$-sites in the $d$--$p$ model.


The present work has been partially supported by the Grant-in-Aid for Scientific Research from the Ministry of Education, Culture, Sports, Science and Technology, Japan.

\end{document}